
\magnification=\magstep1
\hfuzz=4.pt	
%
%
\baselineskip 15 pt
\hsize=15truecm
\vsize=23truecm
\voffset=0pt
\hfuzz=4.pt	
%
\parskip 3pt
\def\simge{\buildrel > \over {_\sim}}

\def\i{\item}

         \def\dh2{$(\nabla h)^2$}

         \def\cl{\centerline}
         \def\c{\centerline}
\def\med{\medskip \noindent}
\def\big{\bigskip \noindent}
\def\ni{\noindent}


\line{ \hfil Preprint HU-TFT-92-53}
\medskip

\cl{\bf Scaling Exponents for Kinetic Roughening in Higher Dimensions}

\bigskip \bigskip

\baselineskip=12pt

\cl{ T. Ala - Nissila$^{1}$, T. Hjelt$^2$,
J. M. Kosterlitz$^3$, and O. Ven\"al\"ainen$^2$}

\medskip
\medskip
\c{$^1$University of Helsinki}
\c{Research Institute for Theoretical Physics}
\c{P.O. Box 9 (Siltavuorenpenger 20 C)}
\c{SF - 00014 University of Helsinki}
\c{Finland}

\medskip
\c{$^2$Tampere University of Technology}
\c{Department of Electrical Engineering}
\c{P. O. Box 692}
\c{SF - 33101 Tampere}
\c{Finland}

\medskip

\cl{$^3$Brown University}
\cl{Department of Physics}
\cl{Box 1843}
\cl{Providence R.I. 02912}
\cl{U.S.A.}

\medskip

\med
\cl{\bf Abstract}

\med
We discuss the results of extensive numerical simulations in order to
estimate the scaling exponents associated with kinetic
roughening in higher dimensions, up to $d=7+1$. To this end,
we study the restricted solid - on - solid growth model, for
which we employ a novel fitting {\it ansatz} for
the spatially averaged height correlation
function $\bar G(t) \sim t^{2\beta}$ to estimate the scaling exponent $\beta$.
Using this method, we present a quantitative determination
of $\beta$ in $d=3+1$ and $4+1$
dimensions. To check the consistency of these results, we also
compute the interface width and determine $\beta$ and $\chi$
from it independently.
Our results are in disagreement with all existing theories and
conjectures, but
in four dimensions they are in good agreement with
recent simulations of Forrest and Tang [{\it Phys. Rev. Lett.} {\bf 64}:1405
(1990)] for a different growth model.
Above five dimensions, we use the time dependence of the
width to obtain lower bound estimates for
$\beta$. Within the accuracy of
our data, we find no indication of an upper critical dimension up to
$d=7+1$.

             \vfill
\line{PACS numbers: \ 05.40, 61.50, 64.60 \hfill}

\med

\eject
\baselineskip=16pt
\med
{\bf 1. INTRODUCTION}

\big
Dynamical behavior of interfaces in random media constitutes
a fundamental problem is statistical mechanics [1], with applications to
e.g. fluid mechanics [2,3], surface growth [4,5], polymer physics [6]
and magnetic flux lines in
superconductors [7]. For a very large class of physical systems, the
essential physics associated with
the dynamics of such objects can be described
in the continuum language by
the Kardar - Parisi - Zhang (KPZ) equation [4,8]:

$${\partial h\over\partial t} = \nu\vec\nabla^2 h +
{\lambda \over 2} (\vec\nabla h)^2 + \eta (\vec r,t) + \mu. \eqno
(1)$$

\ni
In particular, for many simple
ballistic growth models Eq. (1) is now widely accepted
to be the relevant mapping in the continuum limit [1,9]. For these models,
Eq. (1) has a particularly transparent physical interpretation:
the height variable $h(\vec r,t)$ describes a growing interface on
a $d_s=d-1$ dimensional hyperplane, $\mu$ is a constant
driving force and $\nu$ and $\lambda$ describe the effects of
surface tension and lateral growth velocity, respectively.
The random variable $\eta$ is Gaussian and satisfies

$$\langle \eta (\vec r,t) \eta ({\vec r}^{\ \prime},t') \rangle = 2 D
\delta^{d_s}( \vec r - {\vec r}^{\ \prime}) \delta (t-t'), \eqno (2)$$

\ni
with $D$  describing local variations in the deposition
rate.

\med
The properties of the KPZ equation are usually described through the first
two moments of the associated probability distribution function
of the variables
$\{ h(\vec r,t) \}$. Namely, the average interface height $\bar h(t)
\equiv \langle h(\vec r,t) \rangle $ grows
linearly in time, while the surface width $w(L,t)$
which is defined as the standard deviation of heights

$$ w^2(L,t) = \int d \vec r \
 [h(\vec r,t) - \bar h(t)]^2 / L^{d_s} \eqno (3)$$

\ni
displays nontrivial scaling behavior as a function
of time and the linear system size $L$ [4,8,10]

$$w(L,t) \sim L^{\chi} f({t \over L^z}). \eqno (4) $$

\ni
The scaling function $f(x) \propto x^\beta $
for $x \ll 1$, with $\beta \equiv \chi / z$ and
becomes constant for $x \gg 1$. The KPZ equation also satisfies
an exact invariance under a "Galilean" transformation,
which leads to the identity
$\chi + z = 2$ [2,4,8] and leaves only one independent
scaling exponent to be determined.

\med
The scaling indices $\beta$, $\chi$ and $z$ for the KPZ equation
are exactly known in $d=1+1$ dimensions, where the fluctuation - dissipation
relation and Galilean invariance amalgamate to yield $\beta(2) = 1/3$,
$\chi(2) = 1/2$ and $z(2)=3/2$ [1,2,4,8]. These numbers have also
been consistently obtained from direct integrations of the (discretized) KPZ
equation [11,12], and simulations of ballistic growth models [1,9,13-15],
implying universal
behavior as embodied by the KPZ equation.
In addition to the exponents, the scaling
function has recently been calculated numerically using
self consistent mode coupling equations [16]. The concept of universality
between different models has further been strengthened
by recent calculations, where certain amplitude ratios have been
shown to be universal for several ballistic growth models [17-19].

\med
In higher dimensions, where no fluctuation - dissipation
relation exists to fix $\chi$ and where the dynamical renormalization group
efforts at the strong coupling fixed point fail,
exact results are few. Only the weak coupling (ideal interface)
case where $\lambda=0$ is exactly known [20,21], together with
the dynamical exponent $z_c=2$ at the roughening transition
of the KPZ equation to this smooth phase above $d > 2+1$ [22]. Not
surprisingly then, a wealth of approximate analytic schemes [1,23-25] and
numerical simulations [1,9,11,12,14,15,26,27] have been employed
to evaluate the scaling exponents and their behavior in higher dimensions.
However, the situation remains unsettled.
In particular, both recent functional renormalization group calculations
by Halpin - Healy [23] and $1/d$ expansions results of Cook and Derrida [24]
indicate the existence of an {\it upper critical dimension} $d_u$,
above which the strong coupling phase vanishes and $z=2$, $\beta=\chi=0$.
In Halpin - Healy's theory, $d_u=4+1$.

\med
Numerical simulations of various realizations of the KPZ equation
also give contradictory results [1].
Direct attempts to solve the discretized KPZ equations have apparently
been riddled by stability problems, and rather different results have
been reported [11,12]. Neither have the results from simulations of directed
polymers in random media [1,26,27], nor various
ballistic growth models [1,9,14] converged.
One of the most serious recent attempts to determine the scaling
exponents has been the conjecture of Kim {\it et al.} [14],
which was based on numerical results obtained for
the restricted solid - on - solid
growth (GRSOS) model. According to this conjecture,

$$\beta(d)={1\over d+1}, \ \chi(d)={2 \over d+2}, {\rm \ and}
\ z(d)={2(d+1) \over
             (d+2)}, \eqno (5)$$

\ni
which is correct at $d=0+1$ and $1+1$, and implies no finite
upper critical dimension in the problem. However, this result has been
seriously challenged by large scale simulations of Forrest {\it et al.}
[9] on a hypercube stacking model, from which they obtained
$\beta(3)=0.240(1)$ and $\beta(4)=0.180(5)$, which are slightly but
distinctly lower than the predictions of Eq. (5).
To add to the confusion, Kim {\it et al.} [27]
have recently done additional studies of the directed polymer problem,
with results in accordance with Eq. (5) at three and four dimensions.

\med
The purpose of the present work is to try to answer some of the
open questions regarding the KPZ equation in higher dimensions.
The first and the most basic one concerns the actual values
of the scaling exponents. A related question is the controversy
between the results of Forrest {\it et al.} [9], and the conjecture of
Eq. (5). This is particularly important in addressing the question
of universality between ballistic growth models beyond two dimensions.
A third question concerns the existence of an upper critical dimension.
The only numerical work which to our knowledge has addressed this
is the directed polymer simulation of Renz [26], which shows no indication
of $d_u$ up to six dimensions.

\med
To address all these questions, we have undertaken new numerical
simulations of the GRSOS model in higher dimensions, up to
$d=7+1$. In this work, we shall concentrate on dimensions $d \ge 3+1$,
where the original work of Kim {\it et al.} [14] was most seriously
affected by finite size effects, and where possible deviations
from Eq. (5) could be most clearly seen. In particular,
we have developed a new way of extracting
the growth exponents using a {\it fitting ansatz} [15] for the
equal time height
correlation function $G(\vec r,t)$. This fitting ansatz is first
tested in two dimensions, and then applied to extensive numerical
simulations in $d=3+1$, where we have obtained $\beta$ with great
accuracy. Our result, $\beta(4)=0.180(2)$ is well below the prediction
of the conjecture (5), but in remarkably accurate agreement with
the results of Forrest {\it et al.} [9], indicating universality
between these two growth models. The numerical value of $\beta(4)$
is futher supported by standard determinations of $\beta(4)$
and $\chi(4)$ from the surface width $w(L,t)$, for a variety of system sizes.
Furthermore, we have used the surface width in $d=4+1$ to similarly
obtain an estimate for $\beta(5)$ and $\chi(5)$. Above five dimensions,
where finite size effects become prohibitively difficult to overcome,
we have simply used the slope of the growing surface width to obtain
at least lower bound estimates for $\beta$. Since relatively well defined
power laws are found in each case, we find no indication of
$d_u$ up to eight dimensions. Finally, we also discuss briefly
how our fitting
ansatz can be used to study the question of amplitude universality.


\big
{\bf 2. THE GROWTH MODEL}

\big
We define the restricted solid - on - solid growth (GRSOS) model as a ballistic
growth model, where
particles of height unity are randomly deposited on an intially flat
surface [14]. The essential feature of the model is the condition that
for local growth to occur,
the height differences between all the $2d_s$ nearest neighbor columns
(on a hypercubic lattice) must satisfy the condition $\vert \Delta h \vert
\le 1$. If this condition is violated, the deposition attempt is aborted;
however, no other desorption events can take place.
Thus, in the computer simulations each Monte Carlo step (MCS)
consists of $L^{d_s}$ random deposition attempts. We note that neither
parallel nor sublattice growth algorithms have been used in the present study.

\med
Extensive studies of the GRSOS model have
shown [14], that the condition $\vert \Delta h \vert \le 1$
leads to a very rapid approach of the surface width and other
relevant quantities towards nontrivial scaling behavior, which
in two dimensions recovers the exact KPZ exponents.
Moreover, recent explicit numerical
determinations of $\lambda$ in two and three
dimensions [13,17,18], and studies of universal amplitude ratios
in $d=1+1$ [17-19] have further
solidified the relation of the GRSOS model to the KPZ equation.
Furhermore, similarly to the closely related hypercube stacking model [9],
the GRSOS interface can be described in terms of a waiting time
distribution, which corresponds to the discrete directed polymer mapping
of the KPZ equation. Thus, even beyond two dimensions, we expect
the GRSOS model to be able to describe the strong
coupling behavior of the KPZ equation.

\big
{\bf 3. A NEW METHOD OF EXTRACTING THE SCALING EXPONENTS}

\big
As mentioned in the Introduction, there exists a body of numerical
work which has been aimed at the quantitative evaluation of the
scaling exponents. This can in principle done in at least three different
ways: (i) trying to solve a discretized version of the
KPZ equation directly by numerical
iteration, (ii) simulating the directed polymer mapping of the
KPZ equation at zero
temperature, or (iii) simulating a discrete growth model which
belongs to the same universality class as the KPZ equation.
Most attempts to date have been based on (iii) [1]; however
some models are apparently plagued by severe crossover and
finite size effects, which in part
have contributed to the discrepancies reported in the literature.
A related problem in the existing numerical studies lies in
the actual methods used to extract the exponents.
In almost all
previous numerical works [1], $\beta$ and $\chi$ have been determined using
the relations $w(t) \sim t^\beta$, and $w(L) \sim L^\chi$.
However, is has been
shown [14] that the slope of the time dependent
surface width tends to {\it underestimate}
$\beta$, due to a finite size correction. Instead, the
spatially averaged correlation function

$$\bar G (t) \equiv \langle G(\vec r,t) \rangle_r \sim t^{2\beta},
\eqno (6)$$

\med
\ni
where

$$ G (\vec r,t) \equiv \langle [h(\vec x+\vec r,t) -
h(\vec x,t)]^2 \rangle_{\vec x}, \eqno (7)$$

\ni
and the average $\langle
\ \rangle_r$ taken in the {\it asymptotic regime} of $r \gg t^{1/z}$,
has been shown to give
very good estimates of $\beta$ even for relatively small system sizes.
This is due to the fact that $\bar G(t)$ averaged over larger distances
only lacks short wavelength components, which are irrelevant for the
asymptotic behavior. There is an additional advantage of using the
correlation function, because asymptotically

$$G(r,t) \sim \cases{r^{2\chi},  &{\rm for} $r \ll t^{1/z}$ \cr
t^{2\beta}$, &{\rm for} $r \gg t^{1/z}. } \eqno (8) $$

\med
Thus, it is also possible to obtain independent estimates of $\chi$
and $z$ directly from the same set of data, as
we will discuss below.

\med
However, when using $\bar G(t)$ there exists a fundamental problem
in determining the onset of the asymptotic regime, which is to be
used for obtaining $\bar G(t)$. In practice this means introducing a
time dependent cutoff parameter which is somewhat arbitrary.
To overcome this problem, and to fully utilize the information contained
in the correlation function, we have developed a
a novel {\it fitting ansatz} for the correlation function (7) as [15]

$$\hat G(r,t) = a(t)
\{\tanh[b(t)^{1/x} r^{2 \hat \chi (t)/x}]\}^x, \eqno (9)$$

\ni
where $a(t), \ b(t)$ and $\hat \chi (t)$ are fitting parameters, and $x$ is
fixed. This functional form is motivated by the limiting behavior of the
correlation function. Namely, for each fixed time
$1 \ll t \ll L^z$, $\hat G(r,t)
= a(t) b(t) r^{2 \hat \chi (t)}$ for $r \rightarrow 0$, while
$\hat G(r,t) = a(t)$
for $r \rightarrow \infty$. Thus, after fixing $x$,
we can use Eq. (9) to fit {\it the whole function} $G(r,t)$.
First, this gives us

$$\bar G(t) \approx a(t) \sim t^{2 \beta}, \eqno (10)$$

\med
from which $\beta$ can be immediately obtained without a cutoff
parameter. Moreover, the fitting gives us
an independent estimate for $\hat \chi \approx \chi$, as can be seen
from Eq. (8). The third scaling exponent $z \approx \hat z$ can also be
obtained by defining
a radius [14]

$$r_c(t) \sim t^{1/\hat z} \eqno (11)$$

\ni
through the condition that $\hat G(r_c(t),
t) = c a(t) \approx c \bar G(t)$, where $c < 1$ is fixed. By carefully
studying the dependence of $\hat z$ on $c$, it should be possible to rather
accurately estimate $z$, if the ansatz (9) is reliable. This will be
explicitly demonstrated below.

\big
{\bf 4. RESULTS}

\big
{\bf 4.1 d=1+1}

\big
To test the method of fitting the correlation function using the ansatz (9),
we first consider the two dimensional case, for which $\beta(2)=1/3$ and
$\chi(2)=1/2$ are
exact. To this end, we chose to simulate a $L=3000$ system, which is
relatively
small, for which we determined $G(r,t)$ accurately by averaging over 3000
independent runs, up to 800 Monte Carlo steps (MCS) per site.
Standard least squares fitting was then performed to fit to Eq. (9), with
$x=1$. In Fig. 1(a) we show typical results.
The quality of the fits is excellent, and fitting to the $a(t)$ vs.
$t$ curve shown in the inset of Fig. 1(a) between $100 \le t \le 530$ MCS,
we obtain

$$\beta(2) = 0.3324(7). \eqno (12) $$

\med
To compare with the standard method of obtaining $\beta$, we also calculated
the surface width $w(t)$. The analysis of the data is conveniently aided
by defining running exponents, which describe local variations
in the data:

$$ \beta_t \equiv { \log[\bar G(t+n)] - \log [\bar G(t)] \over 2 \log(n)},
   \eqno (13) $$

\ni
where $n$ can be a constant time step, or $n=t$ [9]. The definition for $w(t)$
is completely analogous.
In Fig. 1(b) we show a comparison of the running
exponents between the two quantities. As can clearly be
seen, results from the width indeed tend to underestimate the true
value of $\beta$, giving about 0.327(2). We also tested the effect of
the convolution ansatz [9] to both quantities, which improved results
from the width to $\beta(2)=0.330(3)$. Correspondingly, from
$a(t)$ the result was
$\beta(2)=0.335(2).

\med
To further check the quality of the fitting ansatz, we also monitored the
behavior of the fitting parameter $\hat \chi_t$ as a
function of time, as shown in Fig. 1(c). After an initial transient
there is a plateau which coincides with
the best scaling regime for $a(t)$. Averaging over this region
gives $\hat \chi(2) = 0.498(5)$, which is very close to
the exact value of 1/2. Similarly, estimating $\hat z(2)$
with $c=0.9$ gives $1.503(2)$, and thus $\beta(2) \approx \hat \chi(2)/
\hat z(2) \approx 0.331$. In addition, the Galilean invariance
relation is satisfied very accurately, with

$$\hat \chi(2) + \hat z(2) \approx 2.001. \eqno (14) $$

\med
We note that using the standard methods of
obtaining the scaling exponents, much larger system sizes are needed
for comparable accuracy of the results.

\big
{\bf 4.2 d=3+1}

\big
In four dimensions, we performed extensive simulations of systems of
several sizes: $L=64, 100, 150, 190$ and $250$. We shall first
describe our most accurate
results which are for $L=100$, where both the correlation function
and the width were averaged over 2400 independent runs of up to 400
MCS/site,
with additional test runs up to 800 MCS. In Fig. 2(a) we display results
obtained from using the fitting ansatz with $x=1/2$, while (b) shows
running exponents (from Eq. (13)) from $a(t)$. The quality of the
correlation function fits is again excellent, and a least squares fitting
procedure between $120 \le t \le 350$ MCS gives

$$\beta(4) = 0.180(2), \eqno (15)$$

\med
which is also the result of averaging over the running exponents of Fig.
2(b). Furthermore, the fitting parameter $\hat \chi_t$ in Fig. 2(c)
again displays a plateau, from which $\hat \chi(4) = 0.294(3)$.
To obtain $\hat z(4)$ as well, we carefully checked its dependence
on $c$ for $0.80 \le c \le 0.99$. An average over $c=0.85,0.9$ and
$0.95$ gives 1.709(9), from which the invariance relation becomes

$$\hat \chi(4) + \hat z(4) \approx 2.003. \eqno (16)$$

\med
This high precision check on the Galilean invariance relation
strongly supports the consistency of the fitting ansatz
in these higher dimensions.
However, estimating $\beta(4) \approx
\hat \chi(4)/\hat z(4) \approx 0.172$ demonstrates that $\hat
\chi(4)$ {\it underestimates} the real $\chi(4)$, as we shall directly
show below.

\med
For an independent consistency check of our results, we also determined
$\chi(4)$ directly by calculating the width $w(L) \sim L^{\chi}$ in the {\it
saturated regime} of $t \gg L^z$ for $L=5$, 10, 15, 20, 25, 30, 35, 40,
50 and 60. Averaging over independent configurations was done
until the results seemed to converge, with error bars estimated from variations
between consecutive runs. Results for $w(L)$ are shown in Fig. 3.
Despite considerable efforts, fluctuations in the data for the larger
systems (especially for $L=60$) remained, which demonstrates the difficulty
of a precise determination of $\chi$ with this method. Nevertheless,
for small system sizes we were able to assess the finite size effect.
Namely, $\chi(4)$ starts out near the value of 0.300, increasing with system
size up to about 0.31. Our best estimate from a simple least squares
fit for $20 \le L \le 50$ is

$$\chi(4) \approx 0.308(2), \eqno (17) $$

\med
which, however, does not include finite size effects in a systematic manner.
Thus, the error bar is perhaps not realistic.
Nevertheless, this result is
considerably larger than that obtained for $\hat \chi(4)$,
demonstrating the approximate nature of the fitting ansatz (9).
However, using the direct estimate of $\chi(4)$  above
together with the Galilean invariance relation
yields $\beta(4) \approx 0.182$, which is now fully
consistent with the correlation function data of Eq. (15).

\med
Despite the apparent consistency of the results presented above, we cannot
completely rule out the influence of finite size effects on the value
of $\beta(4)$. To further study this question, we performed additional
simulations for $L=64$ (3000 runs), $L=150$ (75 runs),
$L=190$ (100 runs) and $L=250$ (two runs). The correlation function
fits were performed for $L=64, 150$ and 190;
for $L=250$ only the time dependent
width was computed. Except for the smallest system, for which
the running exponents approach the value of 0.18 from below [15] (see also
Fig. 5), the data from $\bar G(t)$
was simply not good enough for an accurate determination of $\beta(4)$.
Nevertheless, estimating
$a(t)$ between $50 \le t \le 400$ MCS from a fit to the $L=190$ data, gives
$\beta(4) = 0.185 \pm 0.013$, which is consistent with the above results.
However, we can obtain a better grip on
the systematic finite size effects through the direct
dependence of the slope of the width $w(t)$ on $L$. In Fig. 4(a)
we show a comparison of $w(t)$ for all the system sizes studied here,
with running exponents for $L=100$ shown in Fig. 4(b). Using least squares
fits for each case we can summarize the results as:

$$\beta(4) =  \cases{0.173(1), &{\rm for}  \ $L=64$, \cr
                     0.1762(2), &{\rm for} \ $L=100$, \cr
                     0.179(5),  &{\rm for} \ $L=150$, \cr
                     0.180(5),  &{\rm for} \ $L=190$, \cr
                     0.182(1),   &{\rm for} \ $L=250$. \cr} \eqno (18) $$

\med
The details of the fits are as follows:
for $L=64$, $20 \le t \le 150$ MCS; for $L=100$, $50 \le t \le 350$ MCS;
for $L=150$ the result is from two fits for $25 \le t \le 400$ MCS;
for $L=190$ fitting for $50 \le t \le 400$ MCS and $400 \le t \le 840$ MCS
yields $\beta(4)=0.177(1)$ and $0.183(2)$, respectively, the average
of which is 0.180; and finally
for $L=250$, $30 \le t \le 320$ MCS, where the best scaling regime was
found.
Remarkably enough, at least for the range of system sizes studied here, the
scaling exponent $\beta(4)$ seems to saturate around 0.18, which is again
fully consistent with 0.180(2) as obtained
from the $L=100$ system with the fitting ansatz. Thus, our analysis
supports this value as the most accurate estimate of $\beta(4)$.

\med
In analogy to the two dimensional case, we also attempted to use the
convolution ansatz for improving results obtained from the widths.
Rather surprisingly, however,
this did not yield any improvements - e.g. for our best data for
$L=100$, $\beta(4) = 0.17(1)$ was obtained, with rather poor
accuracy. Similary, for $L=190$, $\beta(4) \approx 0.185$. Our
results clearly
indicate that even up to $L=190$, there are additional time dependent
terms present in $w^2(t)$ besides a constant "intrinsic width" [9].

\med
The remarkable feature of the result (15) is that
it is considerably {\it smaller} than that obtained by Kim
{\it et al.} [14], who estimated $\beta(4)=0.20$ using a $64^3$
system. The essential difference between their results and ours lies in
the definition of time, for which they used the average height $\bar h(t)$
instead of the Monte Carlo time $t$.
Namely, it has been shown that there is a finite size correction to
$\bar h(t)$, such that it only becomes directly proportional to time
in the thermodynamic limit [28]. To compare our results directly with Kim
{\it et al.} [14],
we analyzed carefully our data for the $64^3$ system [15]. In Fig. 5
we summarize the results of this analysis, depicting running exponents
from $a(t)$ (again from Eq. (13)) using either MCS or $\bar h(t)$
as a definition of
time. The latter clearly yields a larger value for
$\beta(4) \approx 0.19$ which may explain the
reason for the discrepancy.

\big
{\bf 4.2 d=4+1}

\big
In five dimensions, our analysis parallels that of the four dimensional
case. However, only three system sizes were considered: $L=20$ (3500
runs), $L=50$ (100 runs), and $L=70$ (three runs). The analysis of
the correlation functions proved not to be feasible due to large
fluctuations
in the data. Thus, we determined $\beta(5)$ simply by a least squares
fit to $w(t)$, which is shown in Fig. 6(a). The results are:

$$\beta(5) = \cases{0.124(8), &{\rm for} \ $L=20$, \cr
                    0.135(2), &{\rm for} \ $L=50$, \cr
                    0.139(2), &{\rm for} \ $L=70$. \cr} \eqno (19) $$

\med
Since there is a clear dependence of $\beta(5)$ on system size, we
also analyzed the behavior of the running exponents for $L=50$ and
$70$, which are shown in Fig. 6(b). For $L=70$, the result quoted
above comes from $50 \le t \le 400$ MCS; however, due to a big
fluctuation clearly visible in Fig. 6
somewhat larger values of the running exponents are obtained at the
latest times, e.g. $\beta(5) = 0.150(1)$ for $200 \le t \le 400$ MCS.
Thus we cannot pinpoint the value of
$\beta(5)$ as accurately as for lower dimensions.

\med
As a consistency check, we also calculated the size dependent saturated
width $w(L)$, which is shown in Fig. 7 for $L=5$, 10, 15, 18, 20,
23, 25 and 30. The fluctuations in the data are clearly visible,
but by simply fitting for the six largest systems we estimate

$$\chi(5) \approx 0.245(1). \eqno (20)$$

\med
Using Galilean invariance this gives $\beta(4)
\approx 0.140$, which agrees quite well with the estimate
obtained from the width using
the largest system size $L=70$. However, it is clear that
our analysis here cannot exclude the influence of systematic
finite size effects, which may further increase the value of $\beta(5)$.

\big
{\bf 4.3 d=5+1, 6+1 AND 7+1}

\big
It is already clear from the results of the previous section, that
increasing the substrate dimension further leads to severe finite
size effects. However, we have undertaken an effort to obtain at least
{\it lower bound estimates} for the scaling exponent $\beta(d)$ above five
dimensions. Besides being of fundamental interest in themselves,
these estimates provide a check on the existence of a possible
upper critical dimensionality $d_u$ for the kinetic roughening problem.
Namely, even for relatively small system sizes it should be possible
to distinguish between a power law behavior of the width, as
opposed to logarithmic behavior expected above $d_u$.

\med
The simulations for $d=5+1$, $6+1$ and $7+1$ were done for $L=30$, $17$, and
$11$, respectively, with three independent runs for each case. In
Fig. 8 we summarize the results for $w(t)$. Clear oscillations
due to the layerwise growth are clearly visible
for these relatively small systems. However, in
each case a rather well defined power law behavior can be seen,
and using least squares fitting we obtain the following estimates:

$$\beta(d) \simge \cases{0.107(2), &{\rm for} \ $d=5+1$, \cr
                         0.10(2),  &{\rm for} \ $d=6+1$, \cr
                         0.08(2),  &{\rm for} \ $d=7+1$. \cr} \eqno (21)$$

\med
Our results thus suggest that if a
finite $d_u$ exists it has to be larger than eight.

\big
{\bf 5. SUMMARY AND DISCUSSION}

\big
To summarize, the purpose of this work has been to address three basic
questions regarding the scaling exponents and universality within
kinetic roughening, as described by the strong coupling behavior of the
KPZ equation. First, using a new fitting ansatz method
we have obtained quantitative estimates for
the scaling exponents in four and five dimensions, which disagree with
all existing theories and conjectures. Second, since our most
accurate estimate $\beta(4)
=0.180(2)$ is in excellent agreement with the results of Forrest {\it et
al.} [9]
for a different growth model, these numbers themselves should be universal
(see also Fig. 9).
Why the recent directed polymer simulations of Kim {\it et al.} [27]
differ from these new results, is still an open question.
Third, we have done additional simulations up to eight dimensions finding
no evidence for an upper critical dimension in the problem.

\med
Finally, we would like to briefly discuss the application of the fitting
ansatz (9) to the determination of universal amplitude ratios of the
growth model. The amplitudes are usually defined through the steady state
correlation function [16]

$$ C(\vec r,t) \equiv {\displaystyle \lim_{t_0 \rightarrow \infty} \textstyle}
\langle [\delta h(\vec r+\vec r_0,t+t_0) -
\delta h(\vec r_0,t_0)]^2 \rangle, \eqno (22)$$

\ni
where $\delta h \equiv h - \bar h$, and
for which $C(r=0,t \rightarrow \infty)=B t^{2\beta}$, and $C(r \rightarrow
\infty, t=0) = A r^{2 \chi}$. The universal crossover scale is defined
as the ratio [16]

$$ g^*= {\lambda \over 2} [{A \over B^{z/2}} ] ^{1/ \chi}.
   \eqno (23)$$

\ni
The (non-universal) amplitudes $A$ and $B$, and $\lambda$
have recently been numerically (and analytically) determined for a variety
of models in two dimensions, including the GRSOS model [17-19].
Although Eq. (22)
is {\it not} the correlation function we have calculated in this work,
we can use the ansatz (9) to estimate $A$, by choosing $1 \ll t_0 \ll L^z$
and approximating

$$G(r,t_0) \approx \hat G(r,t_0) \approx a(t_0)b(t_0) r^{2 \chi} \eqno (24)$$

\ni
for $r \ll t_0^{1/z}$. Thus, $A \approx a(t_0)b(t_0)$ assuming the product
$ab$ remains at least approximately constant within the scaling regime.
In two dimensions, we have checked this to be true to a good approximation,
and data for the
$L=3000$ system gives $A \approx 0.71^{+0.04}_{-0.06}$, which is in
reasonably good agreement with the numerical
result $A=0.81$ of Krug {\it et al.} [17] from much larger systems. We note
that for our small system, we expect the finite size effect for the amplitude
to be significant. For reference, we also determined $A(d)$
for $d=3+1$ and $4+1$.
In the former case, $ab$ is roughly independent of system size beyond $L=100$,
which gives $A(4) \approx 0.19^{+0.07}_{-0.04}$, while the single $L=50$
system gives $A(5) \approx 0.21^{+0.04}_{-0.03}$, the latter most likely an
overestimate due to the finite size effect. In both cases, the
product $ab$ slightly decreases as a function of time.
In these higher dimensions,
neither $\lambda$ nor $B$ have been determined. A more detailed discussion
of these amplitudes is beyond the scope of this work, but would be
desirable to further probe the issue of universality between different
models.

\big
{\bf ACKNOWLEDGEMENTS}

\med
We acknowledge the Scientific Computer Centre
of Finland for a generous allocation of computer time, and Mr. Asko
Sainio for his technical help with the Cray XM-P. Prof. Martin
Grant is also acknowledged for fruitful discussions. This research
has been supported by the Academy of Finland.
J.M.K. was supported in part by NSF grant DMR 89-18358 and wishes to
acknowledge
the Research Institute for Theoretical Physics at the University of
Helsinki for its hospitality during his visit.

\vfill \eject

\ni
{\bf Figure Captions:}

\big

\i{Fig. 1} (a) Correlation functions $G(r,t)$ (solid lines) and fitted
           functions $\hat G(r,t)$ for a $L=3000$ two dimensional
           GRSOS model at $t=300,500,700$ and $900$ MCS/site.
           Only part of the functions are shown here.
           Inset shows $\ln [a(t)]$ vs. $\ln (t)$ as obtained from the fits.
           (b) Comparison between running exponents $\beta_t(2)$ of Eq. (13),
           as obtained from $w(t)$ (circles) and $a(t)$ (squares).
           Averaging over these gives $\beta(2)=0.327(2)$ and $0.3337(7)$,
           respectively. Solid line denotes the exact value of 1/3.
           (c) The running fitting parameter $\hat \chi_t(2)$ vs. time.
           Solid line denotes the exact value of $\chi(2)=1/2$.

\i{Fig. 2} $G(r,t)$ (solid lines) and $\hat G(r,t)$
           at $t=90,180,270$ and $360$ MCS/site
           for the $100^3$ GRSOS model. Inset shows
           $\ln [a(t)]$ vs. $\ln (t)$. Only part of the functions are shown.
           (b) Running exponents (Eq. (13)) from $a(t)$, averaging over
           which gives $\beta(4)=0.180(6)$.
           (c) The running parameter $\chi_t(4)$ vs. time.

\i{Fig. 3} Data for the saturated interface width $w(L)$ in four dimensions.

\i{Fig. 4} (a) The surface width for all four dimensional systems studied
               here. The curves have been shifted for clarity.
           (b) Running exponents $\beta_t(4)$ from the width (Eq. (13))
               for $L=100$. Averaging over these
               yields $\beta(4)=0.1762(2)$.

\i{Fig. 5} Running exponents from $a(t)$ (Eq. (13))
           for a $64^3$ system using Monte Carlo time
           (circles) and the average height $\bar h(t)$ as definitions
           of time.

\i{Fig. 6} (a) The surface widths for all three five dimensional systems. The
           curves have been shifted for clarity. (b) Running exponents
           (from Eq. (13)) for $L=50$ (filled circles) and $L=70$.

\i{Fig. 7} Data for the saturated interface width $w(L)$ in five dimensions.

\i{Fig. 8} The time dependence of the
           surface width for systems above five dimensions. Curves
           have been shifted for clarity.

\i{Fig. 9} Scaling of the interface width at $d=3+1$, with $\beta(4)
           =0.180$. System sizes are $L=64$, 100, 150, 190, and 250.
           Scaling is much worse, if $\beta(4)=1/5$ from Eq. (5) is
           used.

\vfill \eject

\ni
{\bf References:}

\big

\i{1.} J. Krug and H. Spohn, in {\it Solids Far From Equilibrium: Growth,
        Morphology and defects}, C. Godreche ed. (Cambridge
        University Press, Cambridge 1991).

\i{2.} D. Forster, D. R. Nelson, and J. M. Stephen, {\it Phys. Rev. A} {\bf
        16}:732 (1977).

\i{3.} J. Koplik and H. Levine, {\it Phys. Rev. B} {\bf 32}:280 (1985).

\i{4.} M. Kardar, G. Parisi, and Y. C. Zhang, {\it Phys. Rev. Lett.} {\bf
        56}:889 (1986).

\i{5.} J. Villain, {\it J. Phys. (Paris) I} {\bf 1}:19 (1991).

\i{6.} M. Kardar and Y-C. Zhang, {\it Phys. Rev. Lett.} {\bf 58}:2087 (1987).

\i{7.} T. Hwa, {\it Phys. Rev. Lett.} {\bf 69}:1552 (1992).

\i{8.} E. Medina, T. Hwa, M. Kardar, and Y-C. Zhang,
        {\it Phys. Rev. A} {\bf 39}:3053 (1989).

\i{9.} B. M. Forrest and L-H. Tang, {\it Phys. Rev. Lett.} {\bf 64}:1405
(1990);
       L-H. Tang, B. M. Forrest, and D. E. Wolf, {\it Phys. Rev. A}
       {\bf 45}:7162 (1992).

\i{10.} F. Family and T. Vicsek, {\it J. Phys. A} {\bf 18}:L57 (1985).

\i{11.} A. Chakrabarti and R. Toral, {\it Phys. Rev. A} {\bf 40}:11419 (1989);
        J. G. Amar and F. Family, {\it Phys. Rev. A} {\bf 41}:3399 (1990);
        K. Moser, D. E. Wolf, and J. Kert\'esz, {\it Physica A} {\bf 178}:215
        (1991); B. M. Forrest and R. Toral, preprint (1992).

\i{12.} H. Guo, B. Grossmann, and M. Grant, {\it Phys. Rev. Lett.} {\bf
64}:1262
     (1990); H. Guo, B. Grossmann, and M. Grant, {\it Phys. Rev. A}
     {\bf 41}:7082 (1990).

\i{13.} J. Krug and H. Spohn, {\it Phys. Rev. Lett.} {\bf 64}:2332 (1990);
     J. M. Kim, T. Ala-Nissila, and J. M. Kosterlitz, {\it Phys. Rev. Lett.}
     {\bf 64}:2333 (1990); D. A. Huse, J. G. Amar, and F. Family,
     {\it Phys. Rev. A} {\bf 41}:7075 (1990).

\i{14.} J. M. Kim and J. M. Kosterlitz, {\it Phys. Rev. Lett.} {\bf 62}:2289
         (1989); J. M. Kim, J. M. Kosterlitz, and T. Ala-Nissila,
         {\it J. Phys. A} {\bf 24}:5569 (1991).

\i{15.} T. Ala-Nissila, T. Hjelt, and J. M. Kosterlitz, {\it Europhys. Lett.}
        {\bf 19} (1):1 (1992).

\i{16.} T. Hwa and E. Frey, {\it Phys. Rev. A} {\bf 44}:R7873 (1991).

\i{17.} J. Krug, P. Meakin, and T. Halpin-Healy, {\it Phys. Rev. A} {\bf
45}:638
     (1992).

\i{18.} J. G. Amar and F. Family, {\it Phys. Rev. A} {\bf 45}:5378 (1992).

\i{19.} L-H. Tang, {\it J. Stat. Phys.} {\bf 67}:819 (1992).

\i{20.} S. F. Edwards and D. R. Wilkinson, {\it Proc. R. Soc. Lond. A}
     {\bf 381}:17 (1982).

\i{21.} T. Hwa, M. Kardar, and M. Paczuski, {\it Phys. Rev. Lett.} {\bf 66}:441
     (1991); S. T. Chui and J. D. Weeks, {\it Phys. Rev. B} {\bf 14}:4978
     (1976); P. Nozieres and F. Gallet, {\it J. Phys. (Paris)} {\bf 48}:353
     (1987).

\i{22.} C. A. Doty and J. M. Kosterlitz, {\it Phys. Rev. Lett.} {\bf 69}:1979
     (1992).

\i{23.} T. Halpin-Healy, {\it Phys. Rev. Lett.} {\bf 62}:442 (1989);
     T. Halpin-Healy, {\it Phys. Rev. A} {\bf 42}:711 (1990).

\i{24.} J. Cook and B. Derrida, {\it Europhys. Lett.} {\bf 10}:195 (1989);
     J. Cook and B. Derrida, {\it J. Phys. A} {\bf 23}:1523 (1990).

\i{25.} H. G. E. Hentschel and F. Family, {\it Phys. Rev. Lett.} {\bf 66}:1982
         (1991).

\i{26.} W. Renz, private comm. (1989); see also Ref. 1.

\i{27.} J. M. Kim, M. A. Moore, and A. J. Bray, {\it Phys. Rev. A} {\bf
44}:2345
     (1991).

\i{28.} J. Krug and P. Meakin, {\it J. Phys. A} {\bf 23}:L987 (1990).

\bye